\documentstyle[epsf]{mn}

\begin{document}
\input epsf

\title[Elliptical galaxies and MOND]{The fundamental plane of elliptical 
galaxies with modified Newtonian dynamics}

\author[R.H. Sanders] {R.H.~Sanders\\Kapteyn Astronomical Institute,
P.O.~Box 800,  9700 AV Groningen, The Netherlands}

 \date{received: ; accepted: }

\maketitle

\begin{abstract}

The modified Newtonian dynamics (MOND), suggested by Milgrom as an 
alternative to dark matter, implies that isothermal spheres with
a fixed anisotropy parameter should
exhibit a near perfect relation between the mass and velocity dispersion 
of the form $M\propto\sigma^4$. This
is consistent with the observed Faber-Jackson relation for elliptical 
galaxies-- a luminosity-velocity dispersion relation with large scatter. 
However, the observable global properties of  elliptical galaxies comprise a
three parameter family;  they lie on a ``fundamental plane'' in a logarithmic
space consisting of central velocity dispersion, effective radius
($r_e$), and
luminosity.  The scatter perpendicular to this plane is significantly less
than that about the Faber-Jackson relation.
I show here that, in order to match the observed properties of 
elliptical galaxies with MOND, models must deviate from being strictly 
isothermal and isotropic; such objects can be 
approximated by high-order polytropic spheres with a radial orbit anisotropy 
in the outer regions. 
MOND imposes boundary conditions on the inner Newtonian regions which 
restrict these models to a dynamical fundamental 
plane of the form $M\propto{\sigma^\alpha}{r_e}^\gamma$ where the exponents
may differ from the Newtonian expectations ($\alpha=2$, $\gamma=1$).  
Scatter about this plane is relatively insensitive to the necessary 
deviations from homology.
\end{abstract}

\begin{keywords}
{galaxies: elliptical-- galaxies: kinematics and dynamics--
 dark matter, gravitation}
\end{keywords}

\section {Introduction}

On a phenomenological level, the most successful alternative to cosmic 
dark matter is the modified Newtonian dynamics (MOND) proposed by Milgrom 
(1983a).  The basic idea is that the deviation from Newtonian gravity or 
dynamics occurs below a fixed acceleration scale-- a proposal which is 
supported in a general way by the fact that discrepancy between 
the classical dynamical
mass and the observable mass in astronomical systems does seem to appear 
at accelerations below $10^{-8}$ cm/s$^2$  
(Sanders 1990, McGaugh 1998).  The fact that this acceleration scale
is comparable to the present value of the Hubble parameter multiplied
by the speed of light ($cH_o$) suggests a cosmological basis
for this phenomenology.

MOND, in a sense, is designed to reproduce flat extended rotation curves
of spiral galaxies and a luminosity-rotation velocity relationship of the
observed form, $L\propto v^4$ (the Tully-Fisher relation).  But apart from
these aspects which are ``built-in", the prescription also successfully 
predicts
the observed form of galaxy rotation curves from the observed distribution
of stars and gas with reasonable values for the mass-to-light ratio
of the stellar component 
(Begeman et al. 1991, Sanders 1996, Sanders \& Verheijen
1998).  Moreover, MOND predicts that the discrepancy between the 
Newtonian dynamical mass and the observed mass should be large in
low surface brightness galaxies-- a prediction subsequently borne out
by observations of these systems (McGaugh \& de Block, 1997, 1998).

The observational success of MOND is most dramatically evident for
spiral galaxies where
often the rotation curve can be a rather precise tracer of the radial 
distribution
of the effective gravitational force;  for hot stellar systems-- elliptical
galaxies-- the predictions are less precise.  Milgrom, in his original papers,
(1983b) pointed out that MOND implies a mass-velocity dispersion
relation for elliptical galaxies of the form $M\propto\sigma^4$.  
If there were no systematic variation of M/L with mass, this,
would become the observed Faber-Jackson relation (Faber \& Jackson
1976).  

In his seminal paper
on this subject, Milgrom (1984) calculated the structure of isothermal
spheres in the context of MOND and drew several very general conclusions:
First, all isothermal spheres, regardless of the degree of anisotropy in
the velocity distribution, have finite mass.  For a one-dimensional velocity
dispersion of 100 to 200 km/s, this mass is inevitably on a galaxy scale.
Second, at large radial distance the density decreases
as $r^{-\delta}$ where $\delta$ is in the vicinity of 4.
Third, there is an absolute maximum on the mean surface density which
is on the order of $a_o/G$ where $a_o$ is the MOND critical acceleration.
For a mass-to-light ratio of three to four in solar units, this would
translate into a surface brightness in the V band of 20 to 
20.5 mag/(arcsec)$^2$ which is characteristic of hot stellar systems
ranging from massive ellipticals to bulges of spiral galaxies to 
globular clusters (Corollo et al. 1997).
Finally, the mass of an isothermal sphere with a specific anisotropy
factor is exactly proportional to $\sigma^4$ where $\sigma$ is the space
velocity dispersion.  All of these conclusions would
apply to  elliptical galaxies, in so far as these objects can be regarded as
isothermal spheres.

Since this work, it has come to light
that the global properties of elliptical galaxies comprise a
three parameter family (Dressler et al. 1987, Djorovski \& Davis 1987);  
that is to say, elliptical galaxies lie on
a surface in a three-dimensional  space defined by the luminosity (L), the 
the central velocity dispersion ($\sigma_o$), and the effective radius
($r_e$); the mean surface brightness ($I_e$)  
may be substituted for either luminosity or effective radius
(i.e., $L=2\pi{I_e}{r_e}^2$).  This surface appears as
a plane on logarithmic plots and, consequently, has been designated as
``the fundamental plane" of elliptical galaxies, with the form
$L\propto \sigma^a{r_e}^b$ where $a\approx 1.5$ and $b\approx 0.8$.  
Because of the small scatter perpendicular to the fundamental plane, 
this three-parameter relationship 
supersedes the Faber-Jackson relation as a distance indicator.  
The usual physical interpretation
of the fundamental plane is that these relations result from the
traditional virial theorem plus a dependence of mass-to-light ratio 
on galaxy mass (van Albada et al. 1995);  although, one must also assume that 
elliptical galaxies comprise a near-homologous family.

It is not immediately clear how the fundamental plane can be interpreted
in terms of modified dynamics, or, indeed, if the fundamental plane is even
consistent with modified dynamics.  The effective virial theorem for
a system deep in the regime of MOND (low internal accelerations) is
of the form $\sigma^4 \propto M$, with no length scale appearing.
On the face of it, this would imply that hot stellar systems should
comprise a two parameter family as suggested by the older Faber-Jackson
relation.

The purpose of this paper is to consider the dynamics of elliptical galaxies
in terms of MOND, particularly with respect to the relations between
global properties.  I demonstrate that the MOND isothermal sphere is
actually not a good representation of high surface brightness 
elliptical galaxies because
the implied average surface densities are too low.  Elliptical galaxies,
within the half-light radius (the effective radius), are essentially
Newtonian systems with accelerations in excess of the critical MOND
acceleration.  This suggests that other possible degrees of freedom
must be exploited to model elliptical galaxies using modified dynamics.

To reproduce the observed properties of high surface-brightness elliptical
galaxies, it is necessary to introduce small deviations from a strictly
isothermal and isotropic velocity field in the outer
regions.  A simple and approximate way of doing this is to consider 
high-order polytropic spheres (all of which are finite in the context of
MOND) with a velocity distribution which varies from
isotropic within a critical radius to highly radial motion in the outer
regions.  The structure of such objects is determined here by
numerical solutions of the hydrostatic equation of stellar dynamics
(the Jeans equation)
modified through the introduction of the MOND formula for the gravitational
acceleration.  I find that
all models characterized by a given value of the polytropic index and
the appropriately scaled anisotropy radius are homologous and exhibit
a perfect mass-velocity dispersion relation of the form
$M\propto \sigma^4$ with no intrinsic scatter.  However a range 
of models over this parameter space is required to reproduce the 
dispersion in the observed
global properties of elliptical galaxies, and strict homology is broken.  
This adds considerable scatter
to the mass-velocity dispersion relation and introduces a third parameter
which is, in effect, the surface density (or effective radius).  Although
these objects are effectively Newtonian in the inner regions, 
MOND imposes boundary conditions which restrict these Newtonian
solutions to a well-defined domain in the three dimensional space of
dynamical parameters-- a dynamical fundamental plane.
Combined with a weak dependence
of M/L on galaxy mass, the fundamental plane in its observed form is 
reproduced.  

Not only the form but also the scaling of the $M-{\sigma_o}-r_e$ relation is
fixed over the relevant domain of parameter space;
this scaling is relatively independent of the detailed 
structure of the stellar system.
Therefore, given the effective radius and velocity dispersion, the mass
of any elliptical galaxy may be calculated and the mass-to-light ratio
can be determined.  For the galaxies in the samples of J{\o}rgensen, Franx,
\& K{\ae}rgaard (1995a,b) and J{\o}rgensen (1999)
the mean M/L turns out to be 3.6 M$_\odot$/L$_\odot$
with about 30\% scatter.  With a weak dependence of M/L with galaxy mass,
the predicted form of the Fundamental Plane agrees with that found by 
J{\o}rgensen et al.

\section{Basic equations and assumptions}

Following Milgrom (1984), I calculate the structure of spherical
systems by integrating the spherically symmetric hydrostatic
equation (the Jeans equation):

$${d\over{dr}}{(\rho{\sigma_r}^2)} + {{2\rho{\sigma_r}^2\beta}\over r}
 = -\rho g \eqno(1) $$
where $\rho$ is the density, $\sigma_r$ is the radial component of the
velocity dispersion, $\beta=1-{{\sigma_t}^2}/{\sigma_r}^2$ is the
anisotropy parameter ($\sigma_t$ is the velocity dispersion in the tangential
direction), and $g$ is the radial gravitational force which, 
in the context of MOND, is given by 
$$g\mu(g/a_o) = {{GM_r}\over r^2} = {{4\pi G}\over{r^2}}{\int_{0}^{r}
{r'}^2 \rho(r')dr'}. \eqno(2) $$ 
Here $M_r$ is the mass within radius r, $a_o$ is the MOND acceleration
parameter (found to be $1.2\times 10^{-8}$ cm/s$^2$ from the rotation
curves of nearby galaxies), and $\mu(x) = x/\sqrt{(1+x^2)}$ is the typically
assumed function interpolating between the Newtonian regime ($x>>1$)
and the MOND regime ($x<<1$).  Because of the assumed spherical 
symmetry, eq.\ 2 is exact in the context of the Lagrangian formulation
of MOND as a modification of Newtonian gravity (Bekenstein \& Milgrom
1984).  However, if viewed as a modification of inertia eq.\ 2 may
only be an approximation in the general case of motion on
non-circular orbits (Milgrom 1994).

There are four unknown 
functions of radius, $\rho$, $\sigma_r$, $\beta$, and $g$;  therefore,
additional assumptions are required to close this set of 2 equations.
First, I take a definite pressure-density relation, that of a 
polytropic equation of state:
$${\sigma_r}^2 = {A_{n\sigma}}\rho^{1\over n} \eqno(3)$$
where $A_{n\sigma}$, for a particular model, 
is a constant which is specified by the given central velocity
dispersion and density, and $n$ is the polytropic index (for an isothermal 
sphere $n$ is infinite).  This is done simply as a convenient way of
providing stellar systems which are somewhat cooler in the outer regions
as is suggested by observations (discussed below).
The anisotropy parameter is further assumed to depend
upon radius as
$$\beta(r) = {{(r/r_a)^2}\over{[1+(r/r_a)^2]}} \eqno(4)$$
where $r_a$ is the assumed anisotropy radius.
This provides a velocity distribution which is isotropic within $r_a$, 
but which
approaches pure radial motion when $r>>r_a$.  Such behavior is typical
of systems which form by dissipationless collapse (van Albada 1982).  
Thus I will be
considering a two-parameter set of models for elliptical galaxies
characterized by a polytropic index $n$ and an anisotropy radius $r_a$.
Additional physical considerations, such as stability, may limit the range
of these parameters.

Milgrom (1984) found that for an isothermal sphere 
with a given constant  $\beta$, there exits a family of MOND solutions
having different asymptotic behavior as $r\rightarrow 0$.
In general, the density approaches a constant value near the center except
for one particular limiting solution where $\rho \rightarrow 1/r^2$.
The global properties (e.g., the mean surface density, the value of 
$M/\sigma^4$, the
asymptotic density distribution at large r) do not vary greatly within 
such a family of solutions, except for models which
are characterized by an unrealistically low central density of
$< 10^{-2}\,M_\odot$/pc$^3$ (see Milgrom 1984, Figs.\ 1 \& 2).  
I find the same to be true
for the high-order polytropic spheres with an anisotropy factor given
by eq.\ 4; i.e., for a given value of $n$ and $r_a$, there is a 
family
of solutions with differing asymptotic behavior at small r.  Again,
because the global parameters do not vary greatly within a family, I consider
below only the limiting solution; i.e., that with the $1/r^2$ density
cusp.  

Before describing the results of the numerical integration of eqs. 1-4,
one important aspect of this system of equations should be emphasized.
The appearance of an additional dimensional constant, $a_o$, in the
equation for the gravitational force, eq.\ 2, provides, when combined
with a characteristic velocity dispersion of the system (the central radial
velocity dispersion, for example), a natural length
and mass scale for objects described by these equations.  This differs from
the case of Newtonian systems where two system parameters (a velocity 
dispersion and central density) are required to define units of mass
and density.  For MOND systems the characteristic
length and mass are given by
$$R_\sigma= {\sigma_r}^2/a_o,\eqno(5a)$$ $$M_\sigma = {\sigma_r}^4/
Ga_o.\eqno(5b)$$  
There is, in addition, a natural scale for surface density given by
$$\Sigma_m = M_\sigma/R_\sigma^2 = a_o/G\eqno(5c)$$ which depends only upon
fundamental constants.  These natural units imply that the properties
of homologous systems described by modified dynamics should scale 
according to these relations; i.e., $M\propto\sigma^4$ with a 
characteristic surface density which is independent of $\sigma$.
However, for the system to be described by modified dynamics it must extend 
into the regime of low accelerations, i.e., where $a<<a_o$.  This is not a
necessary attribute of Newtonian systems which have finite mass and radius 
($n<5$).  Therefore, we would expect that only those higher order 
polytropes ($n>5$ including the isothermal sphere) to be necessarily
characterized by this scaling because the Newtonian solution 
always has infinite extent and mass.  Such polytropes would inevitably
extend to the regime of low accelerations.  

For a velocity dispersion of 200 km/s, typical for 
elliptical galaxies,
we find $R_\sigma \approx 10$ kpc and $M_\sigma \approx 10^{11}$ M$_\odot$-- 
characteristic galactic dimensions.  The manner in which these length
and mass scales arise, in the context of a cosmological setting for the 
dissipationless formation of elliptical galaxies, will
be discussed in future paper.

\section{A test case: the isotropic isothermal sphere}

Eq.\ 1 is numerically solved using a fourth-order Runge-Kutte technique.
The integration proceeds radially outward by 
specifying a central radial velocity dispersion and choosing, 
at a particular radius, the density corresponding to that of the
limiting solution given by Milgrom (1984).  
In solving for the structure of isothermal spheres,
Milgrom used the natural units of length and mass (eq.\ 5) 
to write eq.\ 1 in unitless form.  Here, 
I use physical units: 1 kpc, $10^{11} M_\odot$, and 
1 km/s (in these units $G = 4.32\times 10^5 {\rm (km/s)^2 
{kpc/(10^{11}} M_\odot)}$
and $a_o = 3700 {\rm (km/s)^2/kpc}$.  Although convenient for comparison
with observations, these 
units do obscure the scaling of solutions.

For comparison with Milgrom's results the first systems considered here
are isotropic isothermal spheres ($n$, $r_a$ $\rightarrow \infty $).
The limiting solution of an isothermal sphere with
a specified value of $\beta$ can be scaled as implied
by the natural units (eqs.\ 5).   I verify this by
solving eqs.\ 1-4 for isothermal spheres with radial velocity dispersion 
ranging from 50 km/s to 350 km/s.  For all such spheres, the density
at large radius falls off as $1/r^4$.
This implies, directly from eq.\ 1, that 
$${\sigma_r}^4 = {0.063}{GMa_o}\eqno(6)$$
which is to say, the mass-velocity dispersion relation is exact for MOND
isotropic isothermal spheres.  

Because the density distribution for the limiting solution falls as
$1/r^2$ at small r and as $1/r^4$ in the outer regions, the density profile
of the MOND isotropic isothermal sphere resembles that of the Jaffe
model (Jaffe 1983).  The mean surface density inside the 
projected half-mass radius (the effective radius), $r_e$, is found to be
$$ \Sigma = 0.134 \Sigma_m. \eqno(7)$$
where the characteristic MOND surface density, $\Sigma_m$, 
is given by eq.\ 5c.
Given that $M = 2\pi {r_e}^2\Sigma$, then from eqs.\ 6 and 7 it follows
that $$r_e = 4.36 {\sigma_r}^2/a_o .\eqno(8) $$

For a system with $\sigma_r = 125$ km/s, one would find, 
with these formulae,
that M = $2.4\times 10^{11}$ M$_\odot$ and $r_e = 18$ kpc.  
The surface density distribution of such an object is well-described by
the de Vaucouleurs $r^{1/4}$ law which works well as an empirical
fit to surface brightness profile of elliptical galaxies.

However, the global properties of MOND isotropic isothermal spheres are 
inconsistent with those of high surface-brightness elliptical
galaxies.  Basically, for a given velocity dispersion, the mass is
too large, the effective radius is too large, and the mean surface
density is too small compared to that of actual galaxies.  
\begin{figure}
\epsfxsize=80mm
\epsfbox{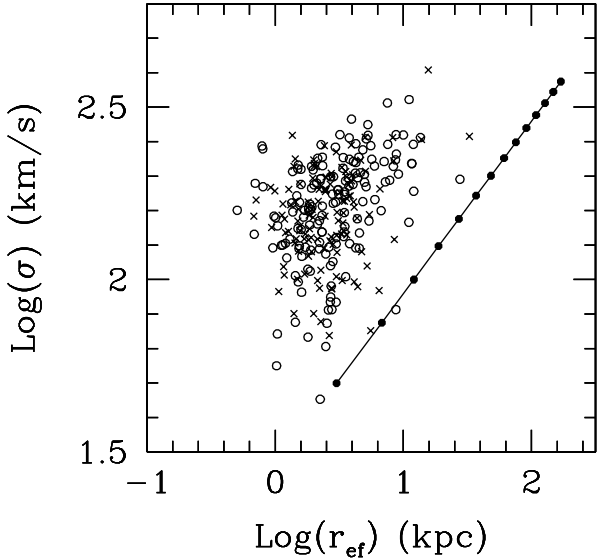}
\caption{A log-log plot of the central velocity dispersion (km/s) vs.
the effective radius (kpc) for cluster elliptical galaxies (open points,
J{\o}rgensen et al. 1995), and for early type galaxies
in the Coma cluster (crosses, J{\o}rgensen 1999). The connected
solid points show the sequence of MOND isotropic isothermal spheres.}
\vfill
\end{figure}
This is evident from Fig.\ 1 which is a logarithmic plot of the 
central velocity dispersion plotted against effective radius 
for a large homogeneously observed sample of galaxies.  The sample 
consists of 154 elliptical galaxies in several clusters (open points)
observed by J{\o}rgensen et al. (1995a,b) combined with 116 early-type
galaxies (crosses) in the Coma cluster (J{\o}rgensen 1999).  The connected
solid points show the sequence of isotropic isothermal spheres (eq.\ 8).  
We see that the distribution of observed
properties is not matched with those of the isothermal spheres; for example,
for galaxies with a central line-of-sight (l.o.s.) velocity dispersion of
about 150 km/s, the effective radii are typically 2.5 kpc, an order
of magnitude lower than that of the MOND isothermal spheres.  Assuming
a mass-to-light ratio of 3 to 4 M$_\odot$/L$_\odot$ 
for the stellar population, the mean observed surface density in actual
elliptical galaxies within 
$r_e$ would be about two or three times larger than $\Sigma_m$; which is to
say, elliptical galaxies 
are well within the Newtonian regime inside an effective radius.
In contrast the mean surface density of MOND isotropic isothermal spheres 
are at least 10 times lower (eq.\ 7) and hence  
fail as representations of luminous high surface brightness elliptical 
galaxies.

The situation becomes worse for constant $\beta > 0$ as is 
evident from the calculation of Milgrom (1984).  Such systems 
are even
deeper in the MOND regime; for $\beta = 0.9$, $\Sigma = 0.0025 \Sigma_m$.  
In order
to represent elliptical galaxies in the context of MOND, we must
allow for the possibility that pressure-supported systems become cooler 
with a radial orbit anisotropy in the outer regions.

\section{Anisotropic polytropes as models for elliptical galaxies}

There is observational evidence that the stellar component of 
elliptical galaxies is not, in general, 
isothermal:  the l.o.s. velocity dispersion is observed to 
decrease with increasing projected radius.  For reasons of stability
such a decrease can probably not be attributed entirely to velocity
field anisotropy (see below).  
Describing this decline by a power law, $\sigma \propto r^{-\epsilon}$,
Franx (1989) finds that typically $\epsilon=0.06$.  A simple
way of introducing this mild deviation from an isothermal state into 
the MOND models is to consider the more general equation of state
expressed by eq.\ 3-- the polytropic gas assumption with large $n$.  

It is straight-forward to demonstrate from eq.\ 1 
that all MOND polytropic spheres of finite $n$ are finite in extent as
well as in mass, unlike the Newtonian case where only polytropes with
$n<5$ have finite extent and mass.  As $n\rightarrow \infty$, the radius 
of the
edge of the sphere also approaches infinity; for high order polytropes
($n>10$) the outer radius is many times larger than the effective
radius.  A polytropic sphere with $n>5$ will always have
a MOND regime which establishes boundary conditions for the inner Newtonian
solution.  Therefore, the polytropic index, which must lie
between 5 and infinity, is a free parameter of such models.

It is unlikely that the stars in elliptical galaxies have a completely
isotropic velocity distribution.  As noted above, a radial 
orbit anisotropy of the form given by eq.\ 4 emerges naturally in 
dissipationless collapse models.  This expression provides a second
dimensionless parameter for characterizing MOND models of elliptical
galaxies:  $\eta=r_a/R_\sigma$,
the anisotropy radius in terms of the characteristic MOND length 
scale

A very general result is that such high order anisotropic polytropic spheres 
have a velocity dispersion-mass relation of the form,
$$ {\sigma_o}^4 = q(n,\eta) GMa_o \eqno(9)$$
where $\sigma_o$ is the {\it central} l.o.s. velocity dispersion.  
That is to say, the ratio of ${\sigma_o}^4$ to $GMa_o$, defined here as $q$, 
depends upon the two dimensionless parameters $n$ and $\eta$. 
This follows from the general scaling relation for systems which extend
into the regime of modified dynamics, eq.\ 5b.
For the pure isotropic isothermal sphere $q_\infty = 0.063$ (eq.\ 6).  
For any given polytropic index
and scaled anisotropy radius, the structure of objects is self-similar over
central radial velocity dispersion, and 
the $M-{\sigma_o}^4$ relation is exact; i.e., such objects form a 
2-parameter family in $\sigma_o$ and $M$.

\begin{figure}
\epsfxsize=80mm 
\epsfbox{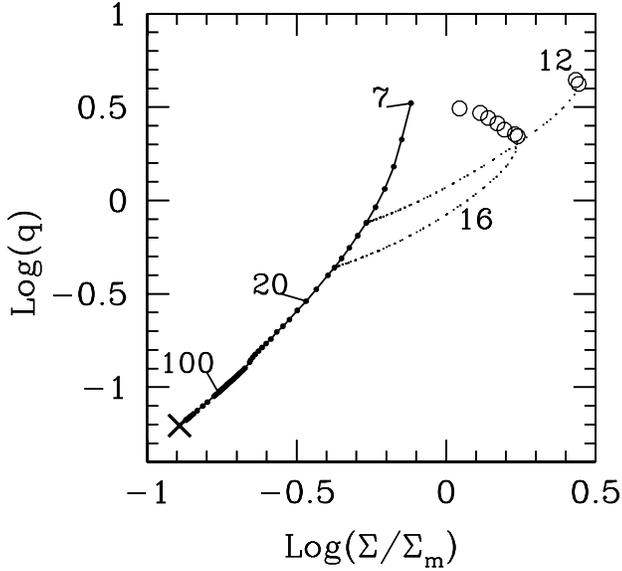}
\caption{A log-log plot of $q$ which is the ratio 
${{\sigma_o}^4}/GMa_o$ ($\sigma_o$ is the central l.o.s. velocity dispersion
and M is the galaxy mass) against the mean surface density within
$r_e$ in terms of the MOND surface density ($a_o/G$) for 
MOND polytropic spheres.  The principal branch shown by the connected
solid points is a sequence of isotropic polytropes; each point is a 
polytropic sphere
of a given index $n$.  The `X' shows the position of the isothermal
sphere and spheres with n=100, 20, and 7 are also indicated.  The parameter
$q$ is the scale factor in the mass-velocity dispersion relation, so each
polytropic sphere has its own $M-{\sigma_o}^4$ relation.  Branching off 
of this curve, for polytropes of n=16 and n=12 are
sequences of anisotropic models where the anisotropy radius, in terms
of the MOND length scale (${\sigma_o}^2/a_o$) decreases from 200 down
to 0.1.  It is seen that increasing anisotropy moves models to regions
of higher surface density.  The open points indicate highly anisotropic
models with $r_a<0.75r_e$.}
\end{figure}

The same is not true if we consider the set of such systems over a 
domain of the two-dimensional parameter space.  In Fig.\ 2 
the connected solid points show the logarithm of
$q$ vs. the logarithm of the mean surface density within the effective 
radius ($\Sigma=M/2\pi r_e^2$) in units
of the MOND critical surface density $\Sigma_m$ for a sequence of 
isotropic polytropes ($\eta = \infty$).  
Each point is a set of models over a range of central $\sigma_o$
but having a specific value of $n$.  The locus of such points define a 
curve with the isothermal
sphere (denoted by an X) at one extreme and, as displayed here, the n=7
polytrope at the other.  
The sequence of isotropic polytropes approaches but does not 
exceed the MOND critical surface density.  Each 
polytrope is self-similar over velocity dispersion
with a perfect mass-velocity dispersion relation,
but its own mass-velocity relation.  The entire set of polytropes 
breaks homology, and for this ensemble of isotropic polytropes, 
$q$ can be considered as a function of the mean
surface density; indeed, over the range of n=30 to n$\rightarrow\infty$,
the curve is well approximated by a power law, i.e.,
$$q = k(\Sigma/\Sigma_m)^\kappa \eqno(10)$$
That is to say, a third parameter, the mean surface density, enters
into the basic dynamical relationship, eq.\ 9.
Substituting eq.\ 10 into eq.\ 9 we find a relation of the form 
$${\sigma_o}^4 = k{{G^{\kappa+1}M^{\kappa+1}}\over{2\pi{r_e}^{2\kappa}
{a_o}^{\kappa-1}}}\eqno(11)$$
This can be viewed as a generalized virial relationship for objects which
extend into the regime of modified dynamics.  For models with a specified
value of $n$ and $\eta$, $\kappa = 0$ and we recover the MOND mass-velocity
dispersion relationship for homologous systems;  models covering a
range of n and $\eta$-- non-homologous models with $\kappa \neq 0$--
comprise a three-parameter family.

If structure of elliptical galaxies could be approximated by
this set of high order isotropic polytropic spheres ($30<n<\infty$), 
then, from eq.\ 11, there would exist a theoretical 
fundamental plane relationship of the form
$$M = K {\sigma_o}^\alpha {r_e}^\gamma \eqno(12a)$$
where $$\alpha = {4\over{\kappa+1}}, \eqno(12b)$$
      $$\gamma = {2\kappa\over{\kappa+1}} \eqno(12c)$$
and   $$K = \Bigl({{2\pi}\over{kG^{\kappa+1}{a_o}^{1-\kappa}}}\Bigr)
             ^{1\over{\kappa+1}}. \eqno(12d)$$
For the sequence of isothermal spheres over this range of $n$, 
$\kappa = 1.5$ which, from eqs.\ 12b and 12c, implies that $\alpha = 1.6$ and
$\gamma = 1.2$.  Thus, in this generalized dynamical relation, the exponents
may differ from the expected Newtonian values ($\alpha=2$, $\gamma=1$).
{\it Significantly, this one dynamical formula (eq.\ 12a) applies to a range
of models which are non-homologous}.

However, pure isotropic polytropic spheres also fail as acceptable
models of elliptical galaxies.  In the models, as in actual ellipticals, 
the l.o.s. velocity 
dispersion declines with increasing projected radius.  
This decline, mild though it is,
is steeper than that typically observed in ellipticals if $n<12$ and 
too shallow if $n>16$.  MOND
polytropes  in the range of n=12 to n=16 exhibit roughly the observed form of
$\sigma(r)$.  For 
polytropes in this range, the mean surface density within the effective 
radius is still significantly lower than that of true elliptical galaxies--
typically by about a factor of 5 assuming a mass-to-light ratio 
of 4 for stellar population of elliptical galaxies.  If such simple 
spherically symmetric models are 
to approximate real elliptical galaxies, it is clear an additional
degree of freedom must enter into the structure equation (eq.\ 1).  
Here we assume that that degree of freedom is provided by the
radial dependence of the anisotropy parameter as represented by eq.\ 4.

In Fig.\ 2 we see the effect of introducing this second parameter, $\eta$,
on the position of polytropes in the log($q$)-log$(\Sigma)$ plane.  
Branching off of the principal curve defined by the sequence of isotropic
polytropes are sequences of models with
$\eta$ ranging from 200 to 0.1 for polytropes of n=12 and n=16.  
When $\eta>> 1$ the models, of course, are very similar
to the MOND isotropic polytropes.
The effect of increasing anisotropy (lower $\eta$) is to increase 
the mean surface 
density of the polytropic spheres-- as is needed to match the observations
of elliptical galaxies.   
The mass and effective radius are decreased but the mean surface density
is higher.

For a given polytropic index, the branch defined by increasing 
anisotropy (decreasing $\eta$) exhibits a maximum
surface density; for n=16 this
maximum is about $2\Sigma_m$ and occurs for $\eta\approx 0.15$;  
that is, for lower $\eta$ (higher anisotropy) the surface brightness
is again lower.  The sequence of anisotropic models is double-valued 
in surface brightness.  These models near the maximum surface density are
quite anisotropic in the sense that the radial orbit anisotropy reaches
within the effective radius;  
all models with $r_a<0.75r_e$ are designated by an open
circle in Fig.\ 2.
The stability of such anisotropic models is questionable (Binney \& Tremaine
1987).

Given the fact that MOND anisotropic polytropes between n=12 and n=16 can
reproduce the approximate decline of the l.o.s. velocity dispersion
with projected radius observed in ellipticals, we can take this as an
observational restriction upon the range of $n$.  The range of the
second parameter,
$\eta$, can also be restricted by excluding all highly anisotropic models
(with $\eta<0.2$) on the basis of possible radial orbit instability.
Fig.\ 3 is the locus of a grid of such models
on the log($q$)-log$(\Sigma)$ plane.  
There are 360 models with $n$ = 12, 13, 14, 15, 16 and 
$\eta$ = 0.2, 0.4, 0.8, 1.6, 3.2, 6.4 and covering a
range of the central $\sigma_r$ between 75 km/s and 350 km/s in steps
of 25 km/s.  

\begin{figure}
\epsfxsize=80mm
\epsfbox{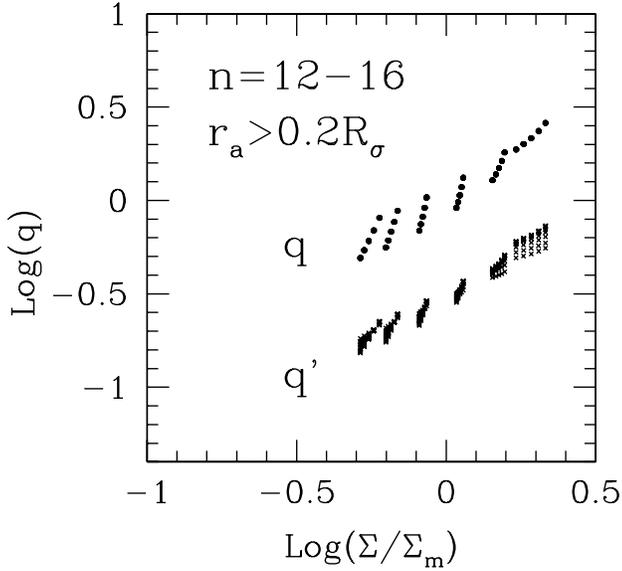}
\caption{The log($q$)-log($\Sigma$) plot for polytropes between n=12
and n=16.  The scaled anisotropy radius is greater than 0.2.
Each point represents a model with specific values of the polytropic
index and scaled anisotropy radius.  There is a perfect
mass-velocity dispersion relation for each point on this plot; although,  
the ensemble of points would exhibit a $M-{\sigma_o}^4$ relation
with considerable scatter since the normalization ($q$) varies by
a factor of five.  Also shown is $q'$ which is the normalization of the
mass-velocity dispersion relation where the intensity-weighted mean
l.o.s. velocity dispersion, $\sigma_d$, within a circular diaphragm of
radius $r_d$ = 0.8 kpc is substituted for the central velocity dispersion; 
i.e., $q'={\sigma_d}^{4-\lambda}{\sigma_m}^\lambda/GMa_o$ with 
$\sigma_m = \sqrt{a_or_d}$.  This is the mass-velocity dispersion relation
which is relevant to actual observations of ellipticals (see text).}
\end{figure}

Each point in Fig.\ 3 represents a particular value of $n$ and
$\eta$ and exhibits its own
perfect $M-{\sigma_o}^4$ relation.  However, the ensemble of 
models is non-homologous
and presents an ensemble of $M-{\sigma_o}^4$ relations.  Because $q$ 
varies by a factor of 5 or 6
this would be the expected intrinsic scatter in the combined
$M-\sigma_o$ relation.
However, over this range of parameter space, the models lie in a 
restricted domain of the log($\Sigma$)-log$(q)$ plane-- 
sufficiently restricted as to define a theoretical
fundamental plane with scatter less than that of the $M-\sigma_o$ 
relation.
A least-square fit to this distribution of points gives $\kappa = 0.98$
in eq.\ 10.  Thus, by eqs.\ 12, this yields a dynamical fundamental plane 
relation near that implied by the Newtonian virial theorem for homologous 
systems, i.e., $\alpha = 2$, $\gamma = 1$.
Because the scatter in $q$ about this power law relation is much less than
the total range in $q$, the scatter perpendicular to the dynamical
fundamental plane is very much reduced.

\begin{figure}
\epsfxsize=80mm
\epsfbox{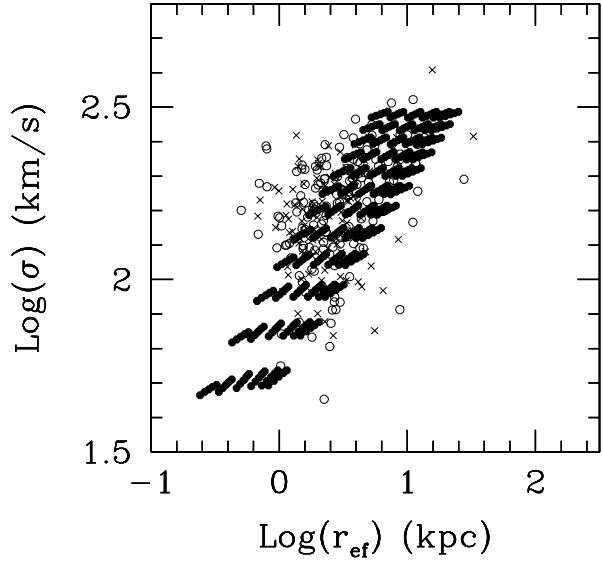}
\caption{The l.o.s. velocity dispersion, $\sigma_d$, within $r_d$
plotted against the effective radius
(as in Fig.\ 1) again for the galaxies from the samples of J{\o}rgensen et al.
but here compared to the ensemble of MOND anisotropic polytropes
(solid points); i.e., the polytropic
index ranges from 12 to 16 with a scaled anisotropy radius greater than
0.2.}
\end{figure}

For comparison with observations, it must be realized that both the 
$M-\sigma$ and fundamental plane relations are altered by
the way in which elliptical galaxies are actually observed.  Specifically,
it is not the velocity dispersion along the very central line-of-sight,
$\sigma_o$, which is measured, but rather the l.o.s. velocity dispersion
, $\sigma_d$, within some finite-size aperture with radius $r_d$.  
The data of J{\o}rgensen et al.
have the advantage that all measured velocity dispersions are 
corrected to a circular aperture with a fixed linear diameter of 1.6 kpc
for H$_o$ = 75.  The appearance of a fixed linear scale has the effect 
of introducing an additional dimensionless
parameter into dynamical relationship, eq.\ 9; i.e., $q$ also becomes
a function of $r_d/R_\sigma$
where $R_\sigma = {\sigma_d}^2/a_o$ is the MOND 
length scale appropriate to the system.  However, this parameter can be
absorbed if it is expressed 
as $r_d/R_\sigma = {\sigma_m}^2/{\sigma_d}^2$ where $\sigma_m = 
\sqrt{a_or_d} = 54.4$ km/s.

When we observe the polytropic models in the same way as real 
galaxies (determining the volume emissivity-weighted l.o.s. velocity 
dispersion in the inner projected 0.8 kpc), the distributions of 
velocity dispersions and effective radii may be compared directly to
the observations of J{\o}rgensen et al.  This is done
in Fig.\ 4 where we see that such models can account for the observed
range in these properties provided that free parameters cover their
allowed ranges: $12\le n \le 16$ and $0.2\le\eta$.  That is to
say, the set of models must be non-homologous in order to explain the 
dispersion in observed properties. 

\begin{figure}
\epsfxsize=85mm
\epsfbox{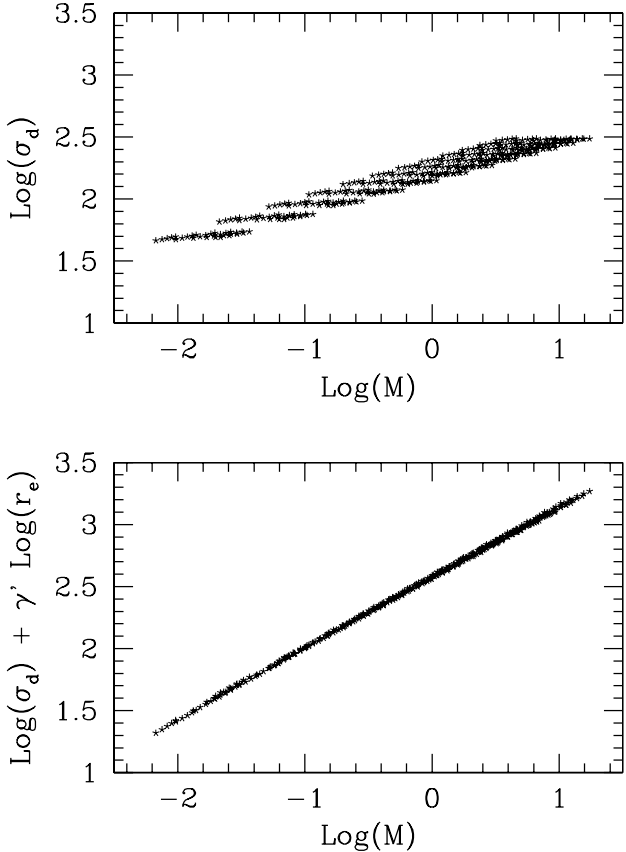}
\caption{The top panel is the mass-velocity dispersion relationship
($M-\sigma_d$)
for the ensemble of anisotropic polytropes.  The polytropic index
ranges between 12 and 16 and the scaled anisotropy radius is greater than
0.2.  This is actually a collection of $M-\sigma_d$ relationships,
one for each combination of the two model parameters.  
The bottom panel shows the
the result of entering a third parameter, i.e., the
best-fit fundamental plane relation.  Here log($\sigma_d$) + $\gamma'$
log($r_e$) is plotted against log(M) and $\gamma'$ is chosen to give
the lowest scatter.  The resulting slope is about 1.76 with $\gamma' = 0.56$,
as implied by the $q'-\Sigma$ relation shown in Fig.\ 3.}
\end{figure}

For these realistically ``observed'' models, the dependence of 
$q$ on $\sigma_d$ is also found to be power law; thus, we may rewrite 
eq.\ 9 as
$${\sigma_d}^4 = q'(\Sigma/\Sigma_m)[\sigma_d/\sigma_m]^\lambda{GMa_o};
\eqno(13)$$  i.e., I explicitly write the dependence of $q$ on $\sigma_d$
leaving the quantity $q'$ as a pure function of surface density. 
In Fig.\ 3 we see that the dependence of $q'$ 
on surface brightness is well-represented by
a power law  with with roughly the same exponent as the $q$ dependence. 
Thus the $M-\sigma$ relation becomes $M \propto {\sigma_d}^{(4-\lambda)}$
and the fundamental plane exponent in eq.\ 12a is  $\alpha = {({4-\lambda})/
({\kappa+1})}$  From the models it is found via least-square fits that 
$\lambda = 0.53$ and $\kappa = 0.98$ implying 
$$M/({10^{11} M_\odot}) = {2\times 10^{-8}} \sigma_d({\rm km/s})^{3.47}
\eqno(14a)$$ for 
the mass-velocity dispersion relation and
$$M/(10^{11}M_\odot)= 3\times 10^{-5} [\sigma_d({\rm km/s})]^{1.76} 
[r_e({\rm kpc})]^{0.98}. \eqno(14b)$$

These $M-\sigma$ and dynamical fundamental plane relations are shown for the
360 models in
Fig.\ 5.  The scatter about the dynamical fundamental plane
is a factor of 10 smaller than that about the $M-\sigma$ relation.

Note in eqs.\ 14 that for this restricted set of models there is a definite 
scaling of both the $M-\sigma$ and the dynamical fundamental plane.  
Using eq.\ 14b to calculate the mass of
the galaxies in the J{\o}rgensen et al. sample, one finds the distribution
of M/L shown in Fig.\ 6 which is a log-log plot of M/L against the 
calculated mass.  Here it is found that $<M/L> = 3.6 \pm 1.2$ 
in solar units (H$_o$ $\approx$ 75) and
and $M/L \propto M^{0.2}$.  This, of course, ignores possibly important
effects such as deviations from spherical symmetry and systematic rotation,
and the fact that real galaxies are almost certainly not characterized by
a pure polytropic velocity dispersion-density relation.  Bearing these 
potential dangers in mind, one could also extrapolate eq.\ 14b down to
globular clusters.  For a sample of globular clusters tabulated by
Trager, Djorgovski \& King (1993) and by Pryor \& Meylen (1993), I find 
that $<M/L> = 1.7 \pm 0.8$ in solar units.

\begin{figure}
\epsfxsize=85mm
\epsfbox{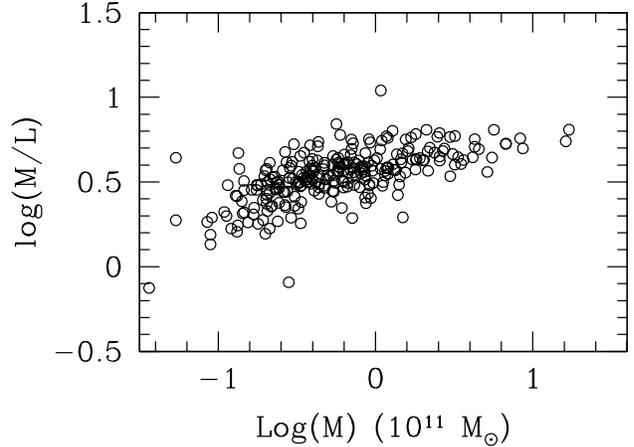}
\caption{The mass-to-light ratio of galaxies in the samples of 
J{\o}rgensen et al. as a function of mass where mass is estimated
from the fundamental plane relation shown in Fig.\ 5 (for the
anisotropic MOND polytropes).  The mean M/L is 3.6 with a 30\% scatter.
As in the strictly Newtonian case there is a weak dependence of
M/L on M.}
\end{figure}

Thus, MOND polytropic spheres in the
range n=12 to n=16 with radial anisotropy beyond the effective radius 
not only reproduce the observed distribution of galaxies in the
$r_e-\sigma_d$ plane but also provide a reasonable
value for the mass-to-light ratio of ellipticals and a weak dependence of
M/L on mass.  As is evident from eq.\ 9, the usual MOND $M\propto 
{\sigma_o}^4$
relation remains, albeit with large scatter due to the necessary deviations
from homology.  Considering the manner in which the central velocity
dispersion is actually measured (within a fixed circular diaphragm) the
relation is altered to $M\propto{\sigma_d}^{3.47}$.  Further, 
considering the necessary dependence of M/L on M, the predicted 
Faber-Jackson relation becomes $L\propto {\sigma_d}^{2.78}$ which is 
consistent with the data of J{\o}rgensen et al.; i.e., a least
square fit to the log(L)-log($\sigma_d$) distribution for this sample of
early-type galaxies yields a slope of $2.6 \pm 0.8$.

The predicted dynamical fundamental plane eq.\ 14b can be converted into the 
more commonly used form by making use of the relation $M=2\pi\Sigma {r_e}^2$; 
then one 
finds $r_e \propto {\sigma_d}^{1.73}{\Sigma}^{0.98}$ where $\Sigma$ is the
mean {\it mass surface density} within $r_e$.  With ${M/L} \propto 
M^{0.17}$ we would then predict a fundamental plane of 
$$r_e \propto {\sigma_d}^{1.23}{I_e}^{-0.84}\eqno(15)$$ where $I_e$ is 
the mean
surface brightness within $r_e$.  Within the uncertainties this is identical
to the fundamental plane defined by the observations of J{\o}rgensen et al.
(1995a,b).

Given the approximations involved in these calculations (primarily the
polytropic gas assumption and the specific radial dependence
of the anisotropy parameter), the actual exponents of the $M-\sigma$ and
fundamental plane relationships are less important than the fact that 
MOND predicts a fundamental plane relation with a scatter which is a 
factor of 10 less than that about the $M-\sigma$ relation.  This is true in
spite of the fact that the set of models must be non-homologous in order to
explain the range of observed properties-- surface density and effective 
radius.  This arises as a natural aspect
of the basic dynamics of systems which extend into the regime of modified
dynamics and need not be accounted for by complicated conspiracies in the
process 
of galaxy formation.  Moreover various mechanisms for the subsequent 
dynamical evolution
of ellipticals (e.g. merging, canabalism) would not effect this relationship.
All that is required is that stellar velocity field in ellipticals
not deviate too dramatically from being isothermal and possess a radial
orbit anisotropy similar to that described by eq.\ 4.  

\section{Conclusions}

The essential results of these calculations can be summarized as follows:
\vskip .2in
\noindent 1.  The dynamics of high surface brightness elliptical galaxies 
span the range from
Newtonian within the effective radius to MOND beyond.  The mean surface
density within $r_e$ is at least twice as large as the MOND
surface density, and the internal accelerations are too large to be within the
domain of modified dynamics. This is consistent with the fact that there is no
large  mass discrepancy or, viewed in terms of dark matter, no need for dark 
matter within the bright inner regions.  However, MOND isothermal spheres
have a mean surface density which is roughly one-tenth the critical surface
density within $r_e$;  they are almost entirely pure MOND objects.
This effectively rules out these objects as models for elliptical galaxies.
\vskip 0.2in
\noindent 2.  In order to reproduce the observed global properties
of high surface brightness elliptical galaxies in the context of
MOND, it is necessary to introduce deviations from a constant velocity
dispersion and strict isotropy of the velocity field in the outer regions.  
This may be done,
in an approximate way, by considering MOND polytropes in the range n=12 
to n=16 
with a radial orbit anisotropy beyond an effective radius ($r_a>0.75 r_e$).
These objects exhibit the mean decline of line-of-sight velocity dispersion 
with projected radius observed in elliptical galaxies.  Moreover, such models
provide reasonable representations of elliptical galaxies with 
respect to the distribution by velocity dispersion and effective radius 
(Fig.\ 4). In order to match these observations, the
models must cover a range in the parameter space of polytropic index and
scaled anisotropy radius which
implies that strict homology is broken.  This breaking of
homology leads to considerable scatter in the mass-velocity dispersion 
relation (and the implied Faber-Jackson relation) while introducing a 
third parameter
which is the mean surface density or effective radius.  The intrinsic 
scatter about this dynamical fundamental
plane is much lower than that about the mass-velocity dispersion
relationship because of the relative insensitivity of this theoretical 
relationship to deviations from homology (Fig.\ 5).  Both the theoretical
$M-\sigma$ and fundamental plane relationships are modified when one 
considers that the central velocity dispersion is actually measured
within a finite size aperture corrected, in the J{\o}rgensen et al. 
observations, to a fixed diameter of 1.6 kpc for all galaxies in their 
samples.
\vskip 0.2in
\noindent 3.  These calculations are highly idealized and apply strictly only
to spherical systems with a perfect polytropic equation of state.  
Moreover, the fact that the models cover 
a range of internal accelerations around $a_o$ means that the detailed
structure is dependent upon the assumed form of the MOND interpolation
function, $\mu(x)$
(eq.\ 2). Nonetheless, when the derived dynamical fundamental plane 
relation is used to estimate the mass of elliptical galaxies from the observed
central velocity dispersion and effective radius, one finds, for the 
galaxies  in the samples of J{\o}rgensen et al., a mean 
mass-to-light ratio of 3.6 M$_\odot$/L$_\odot$ with a dispersion of 30\% 
and a weak dependence
of this M/L on galaxy mass (as in the strictly Newtonian case).  Such
a M/L would seem quite reasonable for the older stellar populations of
elliptical galaxies.  When the predicted dynamical fundamental plane relation
is converted into an observed relationship (on the $r_e$, $\sigma_d$, and
$I_e$ parameter space), the J{\o}rgensen et al. result is recovered
if $M/L\propto M^{0.17}$.
\vskip 0.2in
The principal conclusion is that the existance of a fundamental plane 
with lower intrinsic scatter than that of the Faber-Jackson relation is 
implied by modified dynamics, given 
that high surface brightness elliptical galaxies
cannot be represented by pure MOND isothermal spheres. 
It may be argued that Newtonian dynamics also predicts a fundamental
plane via the traditional virial theorem, and therefore the existance of 
such a relationship in no sense requires modified dynamics.  It
is true that the fundamental plane by itself would not be a sufficient
justification for modified dynamics.  However, a curious aspect of
the Newtonian basis for the fundamental plane is the small scatter about
the observed relation in view of the likely deviations from homology
in actual elliptical galaxies.  The advantage of MOND in this respect
is the existance of a single dynamical relationship (eq.\ 12 or eq.\ 13 ) 
for a range of non-homologous models.  Because an additional dimensional
constant, $a_o$, enters into the structure equation (eq.\ 2), MOND 
self-gravitating objects are more constrained than their Newtonian
counterparts.

In this respect, it is worthwhile to emphasize that a pure Newtonian 
self-gravitating object with a
central line-of-sight velocity dispersion of 200 km/s can have any mass.
But an object with this same velocity dispersion and which extends
at least partially into the regime of modified dynamics can only have a
galaxy-scale mass.  It is the proximity to being
isothermal which requires that elliptical galaxies 
extend into the regime of modified dynamics.
If this one basic requirement is satisfied, MOND inevitably imposes 
boundary conditions
on the inner Newtonian solution-- boundary conditions which restrict 
these objects to lie
on such a well-defined plane in the three dimensional space of 
observed quantities in spite of detailed variations in the structure
between individual objects.
Structural variety does, however, lead to a large intrinsic
scatter in the $M-\sigma$ relation because each distinct
class of objects characterized by an appropriately scaled radial 
dependence of velocity dispersion and degree of anisotropy
exhibits it's own M-$\sigma$ relation with a different 
normalization.  None-the-less, a Faber-Jackson relation does exist
and remains as an imprint of modified dynamics on nearly isothermal 
hot stellar systems. 
\vskip 0.2in
I am grateful to Moti Milgrom and Stacy McGaugh for very useful
comments on the original manuscript.  I thank especially Inger J{\o}rgensen
and Marijn Franx for providing their data on elliptical
galaxies in convenient form and for helpful comments.  I am grateful
to the referee, Massimo Stiavelli, for remarks and criticisms which
greatly improved the content of this paper.

\clearpage
\end{document}